\documentclass[10pt,conference]{IEEEtran}

\usepackage{epsfig}
\usepackage{graphicx}
\usepackage{epstopdf}
\usepackage{amsmath}
\usepackage{amsfonts}
\usepackage{amssymb}

\begin{document}

\title{On Secrecy above Secrecy Capacity}
\author{\IEEEauthorblockN{R Rajesh}
\IEEEauthorblockA{CABS, DRDO\\
Bangalore, India\\
Email: rajesh81r@gmail.com}
\and
\IEEEauthorblockN{Shahid M Shah}
\IEEEauthorblockA{Dept. of ECE\\
Indian Institute of Science\\
Bangalore, India\\
Email: shahid@ece.iisc.ernet.in}\and
\IEEEauthorblockN{Vinod Sharma}
\IEEEauthorblockA{Dept. of ECE\\
Indian Institute of Science\\
Bangalore, India\\
Email: vinod@ece.iisc.ernet.in}}
\vspace{0.01cm}\maketitle

\begin{abstract}
We consider secrecy obtained when one transmits on a Gaussian Wiretap channel above the secrecy capacity. Instead of equivocation, we consider probability of error as the criterion of secrecy. The usual channel codes are considered for transmission. The rates obtained can reach the channel capacity. We show that the "confusion" caused to the Eve when the rate of transmission is above capacity of the Eve's channel is similar to the confusion caused by using the wiretap channel codes used below the secrecy capacity.
\end{abstract}


\section{Introduction}

Information theoretic  security is being widely studied in recent times (\cite{liang}, \cite{blo}). It provides fundamental limits of  secret communication. Unlike  in cryptography,  the information theoretic  techniques, are not based on complexity of computational hardness of certain problems. Furthermore, information theoretic security can also be used in a system along with cryptographic techniques to add additional layers of protection to the information transmission or to achieve key agreement and/or distribution.

Information theoretic formulations for `reliable communication' and `secrecy' were provided in  classic papers of Shannon \cite{shannon} and \cite{shannon1949}. A secret communication system model considers  reliable transmission of information from the transmitter to the intended receiver but this information should not be decoded by the eavesdropper. Shannon characterized reliable communication  using average probability of error and `secrecy' using equivocation. The properties of equivocation are also discussed in \cite{shannon1949}. Shannon considered the case of perfect secrecy when equivocation $H(W|Z^n)=H(W)$, where $W$ is the message transmitted, and $Z^n$ is the received symbol at the eavesdropper. This of course implies $I(W;Z^n)=0$. A weaker form of secrecy is  $I(W,Z^n) \to 0$ as $n \to \infty$ \cite{blo}.  It is further weakened to  $ I(W;Z^n)/n \to 0$ in \cite{maurer}. This implies that the average probability of error to the eavesdropper goes to 1.

 In this set-up of `secret communication' a natural definition of secrecy would be that the intended receiver decodes the message with average probability of error going to zero and the eavesdropper decodes the message with average probability of error going to one. Recently a similar notion of secrecy is also considered in \cite{jean} and \cite{mclaughlin}. In \cite{jean} this notion is used to obtain lattice codes which satisfy such secrecy criteria. In \cite{mclaughlin} it is pointed out that obtaining practical codes satisfying equivocation criterion is very challenging. Therefore they obtain LDPC codes which can provide high \textit{probability of error to Eve} at comparatively low transmir power. We consider this definition in this paper and study the achievable rates on wiretap channel and show that this natural definition leads to improved transmission rates. Then we relate the secrecy obtained via usual codes in this setup to the secrecy obtained via the wiretap codes. We also relate the  probability of error at the eavesdropper to equivocation. In the following we survey the related literature.

`Wiretap channel' introduced and studied by Wyner \cite{wyner1975}, captures the physical communication secrecy problem.  `Wiretap channel' is modelled as a degraded broadcast channel and assumes that the channel between the transmitter and the receiver is better than the channel from the transmitter to the eavesdropper.  Wyner's work was  extended by Leung and Hellman \cite{hellman1976} to the Gaussian channel.   Csisz\`{a}r and K\"{o}rner \cite{csizar1980} considers a general discrete memoryless broadcast channel, and shows that the secrecy capacity is positive if the main channel to the intended user is more capable than of the eavesdropper, and zero if the wiretapper's channel is less noisy.  Practical coding schemes using LDPC codes for wiretap channel are available in \cite{thangaraj}. The secrecy over a fading channel was studied in \cite{gopala2008}. In \cite{bloch2008}, a wire-tap channel with slow fading is  studied where an outage analysis with full CSI of the eavesdropper and imperfect CSI of the eavesdropper was performed. It  is shown that in wireless channels fading helps to provide secrecy rates even if the average SNR of the main channel is poor compared to the eavesdropper's channel.

Our notion of secrecy will be related to equivocation based secrecy via converse results in Information theory. The usual converse result considers average probability of error and is called `weak converse'. It   shows that for a discrete memoryless channel when $R>C$ the probability of decoding error is bounded away from zero \cite{fin}. The strong converse  shows that the maximum probability of decoding error of a MAP (Maximum Aposteriori Probability) decoder tends to one as block length goes to infinity \cite{wolf}. We will use strong converse to formulate the coding schemes. We will also show that the confusion caused by the usual coding schemes to the eavesdropper is not much different from that caused by wiretap codes.

Fano's inequality provides a tight lower bound on the error probability in terms of the conditional entropy. A tight upper bound on the error probability in terms of  conditional entropy is provided in \cite{meh} which also finds relations between probability of error in a MAP decoder and the conditional entropy. Relation between  error probability and conditional entropy are also provided in \cite{ho}. We will use these to get bounds on the equivocation at the Eve for our scheme.

The rest of the paper is organized as follows: In Section II, we define the  model and notation. In Section III, we provide our coding-decoding schemes using average  probability of error as the measure of both reliability and secrecy. We show that using these codes one attains secrecy close to that of the wiretap codes.  Section IV  finds the relation between the chosen criterion of probability of error and equivocation. Section V extends the results to a Gaussian fading channel. We conclude the  paper in Section VI.

\section{Model and Notation}
\begin{figure}
\epsfig{figure=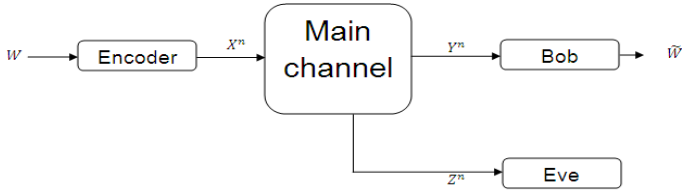,height=2.5cm,width=9cm}
\caption{The Wiretap channel}
\label{fig1}
\end{figure}

We consider a Gaussian Wiretap system (Fig. \ref{fig1}) where a transmitter Alice wants to communicate the message to a legitimate receiver Bob. There is also an eavesdropper, Eve,  who is trying to get access to the message sent to Bob. The transmitter chooses message $W$ for transmission from a set $\mathcal{W} = \{1, 2,..., M\}, M=2^{nR}$, with uniform distribution. These messages are encoded into codewords $(X_{1}, ..., X_{n})$ with a power constraint $E[X^2] \le P$. At time $i$, Alice transmits $X_i$, Bob receives $Y_{i}=X_{i}+N_{1i}$ and the eavesdropper receives $Z_{i}=X_{i}+N_{2i}$. The noise sequences $\{N_{1i}\}$ and $\{N_{2i}\}$ are assumed to be independent of $\{X_i\}$ and also of each other. Also, we assume that $N_{1i} \sim \mathcal{N}(0,\sigma_1^2)$ and $N_{2i} \sim \mathcal{N}(0,\sigma_2^2)$ and $\sigma_1^2 < \sigma_2^2$, where $\mathcal{N}(a,b)$ denotes Gaussian density with mean $a$ and variance $b$.  The decoder at  Bob estimates the transmitted  message as $\tilde{W}$ from $Y^{n} \equiv \{Y_{1}, ..., Y_{n}\}$. It is assumed that Bob as well as Eve know the chosen codebooks. We use $P_e^n(B)$ and $P_e^n(E)$ to denote the average probability of decoding error for block length $n$ at Bob and Eve respectively for the MAP decoder. Our secrecy requirement is that $P_e^n(B) \rightarrow 0$ and $P_e^n(E) \rightarrow 1$.

\section{Capacity Results}

In this section we characterize the achievable rates for the model in Figure \ref{fig1}. Let $C_1= 0.5 \log (1+ P/\sigma_1^2)$ and $C_2=0.5 \log (1+ P/\sigma_2^2)$. Since $\sigma_1^2 <\sigma_2^2$, $C_1 > C_2$. The proof of the following proposition, although simple is given for sake of completeness.

{\bf{Proposition}}: All rates $R$ such that  $0<R<C_1$ are achievable for  a Gaussian wiretap channel such that the probability of decoding  error, $P_e^n(B)$, at Bob goes to zero and the probability of decoding error, $P_e^n(E)$, at eve goes to one as, $n \to \infty$.

{\bf{Proof}}: Consider the region $C_2<R<C_1$. In this region, as in the case of a usual Gaussian channel, we generate $n$ length $iid$ Gaussian codewords with $X \sim \mathcal{N}(0,{P})$. It can be proved using the channel coding theorem for the Gaussian channel \cite{cover} that any rate $R <C$ can be achieved for the Gaussian channel from Alice to Bob. The condition $R <C_1$ ensures reliable communication from Alice to Bob ensuring that $P_e^n(B) \to 0$ exponentially as $n\to \infty$. Also since $R>C_2$, by the strong converse to the coding theorem (in \cite{ari} it is proved for discrete alphabet channels but the result extends to the continuous alphabet channels, see, e.g., \cite{verd}) the average probability of decoding error  for any decoder goes to one exponentially as  $n\to \infty$.
 As this happens for any codebook, it is valid for the chosen codebook that gives reliable communication from Alice to Bob.

Now consider the region $0<R<C_2$. In this region we use coding-decoding schemes as  above but with reduced power $\overline{P} < P$ such that  $0.5 \log (1+ \overline{P}/\sigma_2^2)< R < 0.5 \log (1+ \overline{P}/\sigma_1^2)$. This ensures that we achieve reliable transmission to Bob and secrecy at Eve. ~~~~~~~~~~~~~~~~~~~~~~~~~~~~~~~~~~~~~~~~~~~~~~~~~~~~~~~~~~~~$\blacksquare$

Hence  any rate $0<R<C$ is achievable with Bob getting the message reliably and eve's probability of error as large as we wish.  The block length $n$ should be chosen large enough to satisfy any target probability of error to Bob and Eve. For example suppose we need $P_e^n(B)<\beta _1$ and $P_e^n(E)>\beta _2$. Then for $P_e^n(E)>\beta _2$, rate $R$ and block length $n$ should be such that (\cite{poorverdu})

\begin{equation}
R > C_2 - \sqrt{\frac{V}{n}}Q^{-1}(\beta _2) + \frac{\mathrm{log}n} {2}
\end{equation} 
where $Q$ is the function,

\begin{equation*}
Q(x) = \frac{}{} \int _{x}^{\infty }e^{-\frac{y^2}{2}}dy ,
\end{equation*}
$V$ is the channel dispersion,

\begin{equation}
V = \frac{S_2}{2} \frac{S_2 + 2}{(S_2 +1)^2} \log _2e
\end{equation}
and $S_2=P/\sigma _2^2$, the $SNR$ for Eve. Also, for $P_e^n(B)<\beta _1$, $n$ and $R$ should satisfy the inequality in (Theorem 41, \cite{poorverdu}). These bounds are very accurate even for small $n$. From (1) we see that $R$ does not even need to exceed $C_2$.
This improves the rate over the secrecy capacity definition based on equivocation (\cite{liang}):
\begin{equation}
\label{eqn1}
C=  0.5 \log (1+ P/\sigma_1^2)-0.5 \log (1+ P/\sigma_2^2).
\end{equation}

The equivocation based secrecy systems employ a stochastic encoder whereas in our scheme all rates $ 0 <R < C_1$ can be achieved by the usual encoders employing random  codebooks (in practice one can use LDPC and Turbo codes to obtain rates close to $C_1$). For equivocation based schemes also one can get the required probability of error to Bob and Eve by ensuring large enough block length $n$. However the rate obtained can be quite low and it requires more complex encoders.
\subsection*{Numerical Example}
For $\sigma_1^2=0.1$, $\sigma_2^2=1.5$, and $P=20dB$, $P_e^n(B)$ and $P_e^n(E)$ are plotted in Fig 2 for $n=50,100$ and 200. For $P^n_e(B)$ Gallagers random coding bound and for $P_e^n(E)$ Arimoto's lower bound are plotted (see section IV for these bounds).

\begin{figure}
\epsfig{figure=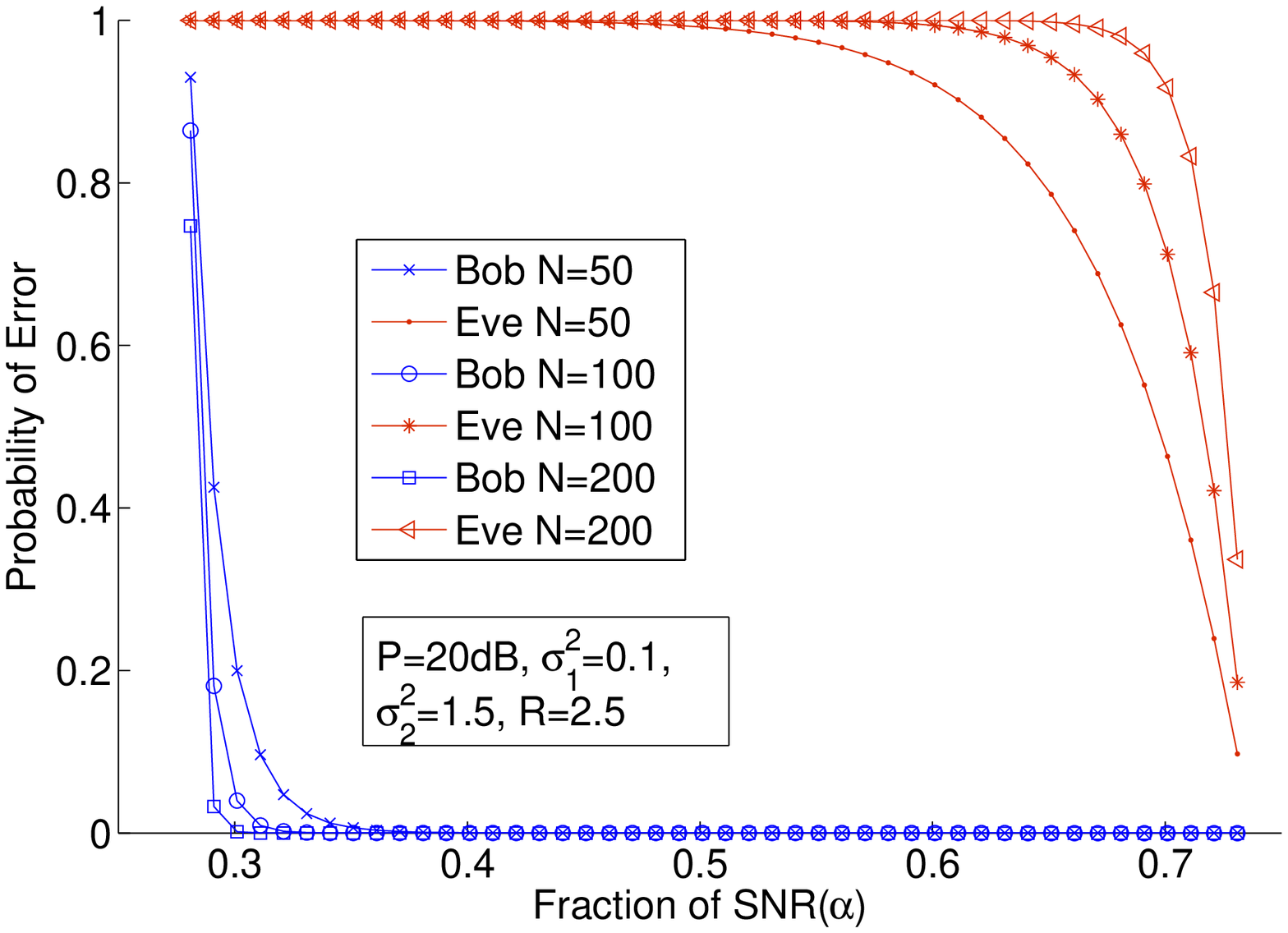,height=8cm,width=17cm}
\caption{Probability of Error vs SNR}
\label{fig2}
\end{figure}

In the following we discuss the capacity achieving encoding and decoding of Theorem 1 and show how it provides security close to the equivocation based schemes. Our arguments are general and can be used for non-Gaussian DMC Wiretap channels also.

To achieve rate $R$, $C_1 >R>C_2$, we select an input distribution $P_X$ such that $I(X;Y)>R>I(X;Z)$ where $X \sim P_X$ and $Y$ and $Z$ are the corresponding channel outputs to Bob and Eve respectively. We generate a codebook with $X_1, ..., X_n$ $iid$ $\sim P_X$, $2^{nR}$ independent codewords, where $n$ is chosen such that Bob's block-probability of error is less than a target value. Bob and Eve can use decoders based on (weak) joint typicality which are asymptotically optimal. Then if Eve receives $Z^n$ on transmission of codeword $X^n$ corresponding to (say) message $1$, probability that $Z^n$ is jointly typical with a codeword corresponding to another codeword is $2^{-nI(X;Z)}$. Let $N$ be the number of codewords other than that of message 1 that are jointly typical with $Z^n$. Let $A_k$ be the event that the $k^{th}$ codeword is jointly typical with $Z^n$. Then

\begin{equation}
N = \sum _{k=2}^{2^{nR}} 1_{\{A_k =1\}}
\end{equation}
 and $\mathsf{E}[N] = (2^{nR}-1)2^{-nI(X;Z)}$.
 
 Since random variable $1_{\{A_k=1\}}$ has an exponential moment, we can also show that $P[|N-\mathsf{E}[N]|>\epsilon]$ decays (super) exponentially with $n$, to zero:
 
 \begin{equation}
 P[\mid N-\mathsf{E}[N]\mid>\delta ]\leq e^{-s\delta} 2^{-2^{nR}(n(R-I(X;Z))-\frac{s}{\mathrm{ln}2}}
 \end{equation}
for any $\delta>0$ and any $s>0$ for all $n$ such that $n(R-I(X;Z))>\frac{s}{\mathrm{ln}2}$. Then, for all $n$, $N+1$(including message $1$ also) will be very close to $2^{n(R-I(X;Z))}$. By symmetry, if Eve has no other information, she should select any one of these codewords. Therefore, probability that Eve selects message 1 $\approx 2^{-n(R-I(X;Z))} $. Since $R$ can be taken close to $C_1$ and $I(X;Z) \leq C_2$, probability of error for the Eve can be $2^{-n(C_1-C_2)}$ which is the maximum rate at which Eve's probability of error decays in the equivocation based secrecy also (when there is \textit{strong} secrecy: $I(X^n;Z^n) \rightarrow 0$). Interpreting secrecy this way is one way of explaining why \textit{strong} secrecy is stronger than \textit{weak} secrecy: $\frac{1}{n}I(X^n;Z^n) \rightarrow 0 $.

For the AWGN Wiretap channel, using sphere packing arguments, one can again connect how the standard coding techniques relate to the secrecy based coding techniques. If the messages are uniformly distributed, ML decoder is optimal. Suppose Eve uses ML decoder. If $x_1(1), ..., x_n(1)$ is transmitted, for ML decoding Eve's decoder decodes as message $\hat{m}$ on receiving $Z^n$ if
\begin{equation}
\hat{m} = arg min \sum _{k=1}^{n} \left(Z_k -x_k (\hat{m})\right)^2.
\end{equation}
This can be reinterpreted as $\sum_{k=1}^{n}(Z_k-x_k(\hat{m}))^2 < r^2$ for $r$ an appropriate constant. Since
\begin{equation}
Z_k - x_k (1) = N_{2k}
\end{equation}
where ${N_{2k}, k\geq 1}$ is $iid \sim \mathcal{N}(0,\sigma _2^2)$ the receiver noise at Eve, $N_{2k}^2$ are $\chi ^2$-central distributed random variables with finite exponential moments in a neighbourhood of $0$. Thus, by (\cite{heath})
\begin{equation}
\sum_{n=1}^{\infty} \beta ^n P\left[\sup_{k\geq n} \mid \frac{1}{k} \sum_{i=1}^k N_{2i}^2 - \sigma _2^2 \mid >\delta\right] <\infty
\end{equation}
for any $\delta >0$ and for some $\beta >1$. In particular $P\left[ |\sum_{i=1}^n N_{2i} - n\sigma_2^2| > n\delta\right] \rightarrow 0$ exponentially. Therefore, for a reasonable probability of error, Eve must have $r>\sqrt{n}\sigma_2$ for even moderate values of $n$. Since again, from the above argument $\sum_{k=1}^n Z_k^2 $ is close to $n(P+\sigma_2^2)$ for any transmitted codeword, maximum number of spheres with radius $\sqrt{n}\sigma_2$, within this sphere is

\begin{equation}
\frac{\left(\sqrt{n(P+\sigma_2^2)}\right)^n}{\sqrt{n\sigma_2^2}} = \left(1+\frac{P}{\sigma_2^2}\right)^{\frac{n}{2}}.
\end{equation}
Since we have $2^{nR}$ codewords, in one sphere of radius $\sqrt{n}\sigma_2$ we will have
\begin{equation}
\frac{2^{nR}}{\left(1+\frac{P}{\sigma_2^2}\right)^{\frac{n}{2}}} = 2^{n(R-C_2)}
\end{equation}
codewords. Thus Eve will confuse with $2^{n(R-C_2)}$ codewords if $r\geq\sqrt{n}\sigma_2$. This is the same number we obtained above if the decoder uses joint typicality.

In equivocation based security Eve is confused by mapping at the stochastic encoder a message into \textit{multiple} codewords. In the above coding scheme, a stochastic encoder \textit{may not} be used, but still the decoder at Eve is \textit{confused} among multiple codewords. The only difference is that unlike in equivocation case the codewords now with which Eve confuses represent \textit{real messages} (i.e, carry useful information) for Bob thus increasing the transmission rate for Bob.   Furthermore, importantly, since in equivocation based approach the total number of messages to be sent is $2^{nR} < 2^{n(C_1-C_2)}$, Eve confuses among $2^{nR}$ messages only, which is no more than the number we obtained in our approach. Thus in a way, the secrecy(confusion) our approach is providing is \textit{no less} than in the equivocation case although we can transmit at rates upto $C_1$. 

It will perhaps be useful in our  setup that if two messages $W_1$ and $W_2$ are 'close' to each other in some meaningful way then they should be coded into $X^n$ sequences which are not close to each other in Euclidean space (for AWGN channel, similar care can be taken for other channels). Actually it seems to be the typical case for codes with rates close to capacity.

Wyner \cite{wyner1975} also discusses the rate region $(R,R_e)$ where $R_e = \limsup_{n \rightarrow \infty}\frac{H(W|Z^n)}{n}$. These Rates in fact turn out to be $R\leq C_1$ and $R_e \leq C_1 - C_2$. Our codebook using the usual point to point capacity achieving code also meets these criteria. But considering secrecy from probability of error point of view advocates using rate point $(C_1,C_1-C_2)$ (in fact sending at rate upto $C_1$ in which \textit{all messages} are secure from Eve) unlike the equivocation view where $(C_1-C_2,C_1-C_2)$ is the preferred point (\cite{liang}, \cite{elgamal}).

\section{Relation between probability of error and Equivocation}

In order to further relate our coding scheme to equivocation based secrecy, in this section we compute equivocation at Bob and Eve for our coding. 

When $C_2<R<C_1$, $P_e^n(B) \rightarrow 0$, exponentially and $P_e^n(E) \rightarrow 1$ exponentially. By Fano's inequality \cite{cover} 

\begin{equation}
\label{eqn2}
H(W|Y^n) \le H(P_e^n(B)) + P_e^n(B) nR 
\end{equation}
where $H(p) = -p\log (p) -(1-p)\log (1-p)$. If $P_e^n(B) \leq e^{-\alpha n}$ then, expanding $H(p)$ at $p=0$ via Taylor series we get $H(p)=p^{\prime}\log ((1-p^{\prime})/p^{\prime})$ for some $p^{\prime} \in (0,p)$ and hence $H(P_e^n(B) \leq \alpha ne^{-\alpha n}$. Thus $H(W|Y^n)$ decays exponentially too. Similarly, the upper bound (\ref{eqn2}) for $H(W\mid Z^n)$ goes to $nR$ exponentially.

From \cite{meh} we get more refined results. For Bob, for any $0\leq \rho \leq 1$,
\begin{equation}
\label{bob_exp}
H(W\mid Y^n) \leq \left(1+\frac{1}{\rho}\right)e^{-n[E_0(\rho,p)-\rho R]}
\end{equation}
where $p$ is the distribution of $X$, $E_0(\rho,p)$ is the Gallager's random coding exponent (\cite{gal}),

\begin{equation}
\label{gall_bnd}
E_0(\rho,p)=-\mathrm{log}\left( \int_y \left( \int_x p(x)p(y\mid x)^{\frac{1}{1+\rho}}dx \right)^{1+\rho}dy\right)
\end{equation}
and $p(y\mid x)$ is the channel transition function. Thus, $H(W\mid Y^n) \rightarrow 0$ exponentially.

For Eve, since $I(W;Z^n)\leq nC_2$, $\frac{1}{n}H(W\mid Z^n) \geq R-C_2$ for all codebooks.

A more accurate lower bound on $H(W\mid Z^n)$ can be obtained from \cite{meh}, Theorem 1,

\begin{equation}
\label{merhav_lb}
H(W\mid Z^n)\geq \phi ^*(\pi (W\mid Z^n)),
\end{equation}
where $\phi^*$ is a piecewise linear, continuous, non-decreasing, convex function provided in \cite{meh} and $\pi (W\mid Z^n)$ is the average probability of error for the MAP decoder at Eve. $\pi (W\mid Z^n)$ in (\ref{merhav_lb}) can be replaced by the Arimoto's lower bound (\cite{ari})

\begin{equation}
\label{prob_err_lb}
\pi(W\mid Z^n) \geq 1-e^{-n\left(E_0(\rho,p)-\rho R\right)}, ~~~~0\geq \rho \geq -1.
\end{equation}
Minimizing the exponent with respect to $p$ and maximizing w.r.t $\rho$ provides a universal lower bound.

 
\section{Fading channel}

A fading wiretap channel model can be mathematically represented as:
\begin{equation}
\label{model1}
Y_{i}= {h}_{i}X_{i} + N_{1i},
\end{equation}
\begin{equation}
\label{model2}
Z_{i}={g}_{i}X_{i} +   N_{2i},
\end{equation}
where ${h}_{i}, ~{g}_{i}$ are the normalized complex channel gains from Alice to Bob and Eve respectively at instant $i$. We assume that these gains are available at Alice, Bob and Eve. We take $\sigma_1^2=\sigma_2^2=1$. Also denote the instantaneous power gains by $q(i)= |h(i)|^2$ and $r(i)= |g(i)|^2$. The rest of the model is same as in Section II.  For this case the capacity achieving signalling scheme based on equivocation at Eve is  to transmit at instances when  $q > r$ and adapt the instantaneous power according to $q$ and $r$. The capacity is \cite{gopala2008}:
\begin{eqnarray}
\label{eqn9}
C_s&=& \int_0^\infty\int_{r}^\infty[ \log(1+q P^*(q, r))-\nonumber\\
&&~~~~~~~~~\log(1+rP^*(q, r))]f(q)f(r)dq dr
\end{eqnarray}
where $E[P^*(q,r)] = P$, 
\begin{eqnarray}
\label{eqn10}
P^*(q,r) &=& 0.5 [ \sqrt{(1/r-1/q)^2+ 4/\lambda(1/r-1/q)}+ \nonumber \\
&&~~~~~~~~~~~~~~(1/r+1/q)]^{+}
\end{eqnarray}
and $\lambda$ is chosen to  satisfy the power constraint with equality.

Now we extend our proposition to include fading. We use a scheme similar to that in  \cite{gopala2008} but with the condition $P_e^n(B) \to 0$ and $P_e^n(E) \to 1$ as $n \to \infty$. We can make the following claim: All rates $R$ such that  $0<R<C_1$ are achievable for  a Gaussian wiretap channel such that the probability of decoding  error, $P_e^n(B)$, at Bob goes to zero and the probability of decoding error, $P_e^n(E)$, at eve goes to one as, $n \to \infty$ where 

\begin{equation}
\label{eqn11}
C_1= \sup_{P(q,r)} \int_0^\infty\int_{r}^\infty[\log(1+q P(q, r))]f(q)f(r)dq dr.
\end{equation}.The optimal power allocation is 'water-filling' w.r.t. the distribution of $q$ conditioned on $q>r$ (i.e., we transmit only when $q>r$ and optimize power for this case).

\section{Conclusions}
This paper uses a new notion of secrecy for improving the rates in a Gaussian wire-tap channel. This new notion is based on using probability of error as the measure of secrecy at both the intended receiver and also the eavesdropper. In such a set up, it is shown that  the random codes used for point to point communication can be used to provide secrecy as well. We relate this notion of secrecy to the equivocation based secrecy. The results are also extended to a fading channel.

We believe that this notion of secrecy is strong enough for practical purposes (and in a way can provide secrecy equivalent to the equivocation based approach) , but uses usual channel codes and provides the maximum possible rates for reliable transmission to the intended receiver. Although we have shown these concepts on a Wiretap channel, these ideas can obviously be used on other channels as well.


\end{document}